\documentclass{pasj00}
\Received{2008 June 12}
\Accepted{2008 August 9}
\Published{$\langle$publication date$\rangle$}
\SetRunningHead{Uchiyama, H. et al.}{Suzaku observation of HESSJ1825$-$137}
\usepackage{times}
\begin{document}
\title{Suzaku Observation of HESS~J1825$-$137: \\ Discovery of Largely-Extended X-rays near from PSR~J1826$-$1334}
\author{Hideki \textsc{Uchiyama}, Hironori \textsc{Matsumoto}, Takeshi Go \textsc{Tsuru}, and Katsuji \textsc{Koyama} }
\affil{Department of Physics, Graduate School of Science, Kyoto University, Sakyo-ku, Kyoto 606-8502}
\email{uchiyama@cr.scphys.kyoto-u.ac.jp}
\and
\author{Aya \textsc{Bamba}}
\affil{Institute of Space and Astronautical Science, Japan Aerospace Exploration Agency, \\
3-1-1 Yoshinodai, Sagamihara, Kanagawa 229-8510}

\maketitle

\begin{abstract}
We observed the brightest part of HESS J1825$-$137 with the
Suzaku XIS, and found diffuse X-rays extending  at
least up to $\timeform{15'}$ ($\sim 17$~pc) from the pulsar
PSR~J1826$-$1334. 
The spectra have no emission line, and are fitted with an absorbed power-law model. 
The X-rays, therefore, are likely due to synchrotron emission from a pulsar wind nebula.  
The photon index near at the pulsar ($r\leq1.5'$) is  $\Gamma=1.7$
while those in $r=1.5-16'$ are nearly constant at $\Gamma=2.0$.
The spectral energy distribution of the Suzaku and
H.E.S.S. results are naturally explained by a combined process; 
synchrotron X-rays and $\gamma$-rays by the inverse 
Compton of the cosmic microwave photons by  high-energy electrons 
in a magnetic field of $\sim 7~\mu$G.  
If the electrons are accelerated at the pulsar, the electrons must be 
transported over 17~pc in the synchrotron life time of 1900~yr, with a velocity of 
$\geq 8.8 \times 10^3$~km~s$^{-1}$. 
\end{abstract}

\section{Introduction}
PSR~J1826$-$1334 was discovered by the Jodrell Bank radio survey. 
The spin period ($P$) is 101~ms and the period derivative ($\dot{P}$) is
$7.5\times10^{-14}~\mathrm{s~s^{-1}}$, implying a spin-down
luminosity ($\dot{E}$) of $2.8 \times 10^{36}
\mathrm{~erg~s^{-1}}$ and a characteristic age ($\tau$) of
21~kyr~\citep{Clifton1992}. The distance to
PSR~J1826$-$1334 has been measured as $3.9 \pm 0.1$~kpc by
the dispersion measure of its radio pulses~\citep{Taylor1993,Cordes2002}. 

From the spin-down luminosity and characteristic age, PSR~J1826$-$1334
may be a member of the Vela-like pulsars, in the evolutionary stage of 
the high-energy emission processes dominance. There are pulsed non-thermal
and blackbody emission from a pulsar, an extended nebulosity powered by pulsar
wind  (Pulsar Wind Nebula; PWN), and thermal and/or non-thermal emission 
from a supernova remnant (SNR)~\citep{Bec97,Gaensler2003PSRB1823}.

Extended X-rays around PSR~J1826$-$1334 had been
found with the ROSAT~\citep{Finley1996} and ASCA~\citep{Sakurai2001} 
observations, possibly due to PWN, but no extended radio emission was 
found ~\citep{Bra89,Gae00}.

XMM-Newton resolved the extended X-rays into two parts; "core",
a bright and elongated emission in the east-west direction with an extent of
\timeform{30"}, and  "diffuse", a larger scale with lower surface 
brightness emission extending by $\sim \timeform{5'}$  mostly to the south of 
the pulsar \citep{Gaensler2003PSRB1823}. 
Spectra of both components can be fitted with an
absorbed power-law model, and hence are suggested to be a synchrotron origin.
The photon index of the core  and diffuse components of  $\Gamma \sim 1.6$ and  
harder $\Gamma$ of $\sim 2$, respectively.
The absorbing column densities $N_{\mathrm{H}}$ are both about 
$1 \times 10^{22}~\mathrm{cm^{-2}}$.
\citet{Gaensler2003PSRB1823} interpreted both the components as 
parts of a single PWN and designated as G18.0$-$0.7.
They argued that the non-detection
of the PWN in the radio band is due to the limited  sensitivity 
for a nebula of this large size.  The one-sided
morphology of the diffuse component suggests that the PWN is
extended exclusively to the south of the pulsar. 

The Chandra observation~\citep{Pavlov2008} confirmed the existence of
the two components, and measured the X-ray spectrum of the
pulsar itself for the first time. The photon index of the
pulsar is $\Gamma \sim 2.4$, and the luminosity is $8 \times
10^{31}~\mathrm{erg~s^{-1}}$ in the 0.5--8 keV band.

HESS~J1825$-$137 is one of the handful 
Very High Energy (VHE) $\gamma$-ray sources on the Galactic plane discovered during the H.E.S.S. Galactic plane survey~\citep{Aharonian2005a, Aharonian2005b, Aharonian2006Survey}. It is extending to the south of the radio pulsar PSR~J1826$-$1334. 
Subsequently, deep follow-up observation was made with H.E.S.S.
~\citep{Aharonian2006HESSJ1825}

The $\gamma$-ray emission is largely extended to $\sim \timeform{1D}$,
i.e. $\sim 70$~pc in size at the distance of 4~kpc. It has a
larger scale one-sided morphology than X-rays toward the south of the pulsar. 
However, the detailed morphology is different from the X-ray;
the surface brightness decreases and $\gamma$-ray  spectrum shows  softening
with the distance from the pulsar.  These suggest that the HESS~J1825$-$137 is 
a PWN powered by PSR~J1826$-$1334. 

The most probable scenario for the VHE $\gamma$-ray emission is
inverse Compton (IC) scattering of the cosmic microwave background (CMB) photons 
with ultra-relativistic electrons accelerated by the pulsar, which is
also responsible for the X-ray emission via the synchrotron process.
However, the VHE $\gamma$-rays are more extended than
the X-rays. Furthermore, the X-rays are weak at the peak position 
of the $\gamma$-rays.  
If the $\gamma$-rays have the electron origin (IC), the typical energy of
electrons is $\sim$20~TeV. On the other hand, the typical
energy of electrons emitting synchrotron X-rays is
$\sim$100~TeV assuming a magnetic field of $B=10~\mu$G. 
These apparent differences between X-rays and $\gamma$-rays may play
a key role for the PWNe physics, such as the energy transport
mechanism of and the history of energy injections to electrons. 

The extent of the diffuse X-ray emission, however,
may be underestimated, due mainly to the limited sensitivity XMM-Newton and 
Chandra for diffuse X-rays.  We hence observed the PWN with the X-ray 
Imaging Spectrometers~\citep{Koyama2007XIS} on board
the Suzaku satellite~\citep{Mitsuda2007Suzaku}.

In this paper, we will introduce the observations in section
2, our analysis and results are explained in section 3, and
we will discuss our results in section 4.  Uncertainties are
quoted at the 90\% confidence range unless otherwise stated.


\section{Observations and Data Reduction}
\begin{figure}
\begin{center}
\FigureFile(80mm,50mm){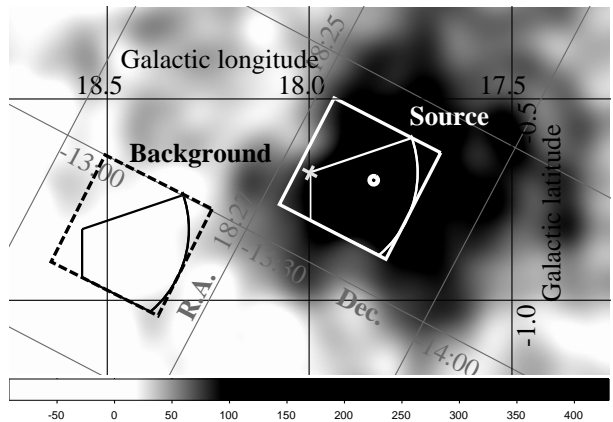}
\end{center}
\caption{
Suzaku field of view overlaid on the H.E.S.S. smoothed
excess map, which is the same as figure 1 in
\citet{Aharonian2006HESSJ1825}. The scale is in units of
integrated excess counts. We also show the position of
PSR~J1826$-$1334 (cross mark), the peak of the VHE
$\gamma$-ray emission (white circle) and the regions from 
which we made radial profiles (thin line) in section~\ref{ch:image}
\label{fig:obspos}}
\end{figure}

We observed PSR~J1826$-$1334 on 2006 October 17--19, with the field center 
at the VHE $\gamma$-ray peak of HESS~J1825$-$137 (source region), and an offset position on 2006
October 19--20, where no VHE ${\gamma}$-ray has been
detected (background region).  The Galactic longitude of the background region
was selected to be almost the same as that of the source
region, so that the Galactic Ridge X-ray Emission (GRXE) at
the source position can be reliably
subtracted~(e.g. \cite{Wor82,War85,Koy86,Yam93,Yam96,Kan97,Sug01,Tan02,
Ebisawa2005,Rev06a,Ebisawa2008}).  The two observations are
summarized in figure~\ref{fig:obspos}\footnote{We used the
image fits file of the H.E.S.S. website at
http://www.mpi-hd.mpg.de/hfm/HESS/public/publications/HESSJ1825II\_Fig1.fits.} 
and table~\ref{tab:observation}.

\begin{table*}
\begin{center}
\caption{Log of Suzaku observations.\label{tab:observation}}
\begin{tabular}{ccccc}	\hline \hline
Region	& Target coordinate (R.A., Dec.)\footnotemark[$*$]	& Obs. start time 			&	Obs. end time	& Effective exp. (ks) \\ \hline
Source & $(\timeform{18h26m00s}, \timeform{-13d41m42s})$ & 2006/10/17 19:37:16   & 2006/10/19 04:02:15	&50.3 \\
Background & $(\timeform{18h27m36s}, \timeform{-13d15m36s})$	& 2006/10/19 04:03:16  & 2006/10/20 12:10:25    &52.1 \\ \hline
\end{tabular}
\end{center}
\hspace{2em}*Equinox in J2000.
\end{table*}

The observations were made with the X-ray Imaging Spectrometer (XIS)
on the focal planes of four X-ray telescopes (XRT) onboard Suzaku. 
The XIS system consists of four CCD detectors (XIS~0, 1, 2, and 3).
XIS~0, 2, and 3 have front-illuminated (FI) CCDs, while XIS~1 contains a
back-illuminated (BI) CCD.  A combination of  the XIS and the XRT provides
a spatial resolution of $\sim\timeform{2\prime}$ in a half-power diameter (HPD) 
and a field-of-view (FOV) of $\timeform{18\prime} \times
\timeform{18\prime}$.  Using the calibration sources ($^{55}$Fe), 
we confirmed that the energy resolution of the XIS was $\sim 210$~eV at 5.9~keV. 
The details of Suzaku, the XIS, and the XRT are given in
\citet{Mitsuda2007Suzaku}, \citet{Koyama2007XIS}, and
\citet{Serlemitsos2007XRT}, respectively.

The XIS was operated in the normal clocking full window
mode and the data of the two editing modes, $3 \times 3$ and $5 \times 5$, 
were combined. 
We used the processed data version 2.0.6.13 and the HEADAS
software version 6.4. After removing the epoch of low-Earth
elevation angles less than 5 degrees (ELV $<5^{\circ}$), the
day earth elevation angle less than 20 degrees (DYE{\_}ELV
$<20^{\circ}$) and the South Atlantic Anomaly (SAA), the
effective exposure times were about 50.3 ks and 52.1 ks for
the source and background regions, respectively. Finally we
removed flickering pixels and hot pixels using {\tt cleansis}.

\section{Analysis and Results}
\subsection{Background Region\label{ch:BG}}

In the background region, 
we found four point sources. The brightest source at $(\alpha,
\delta)_{\rm J2000}=$ ($\timeform{18h27m32s}$,
$\timeform{-13D17m41s}$) is recorded as a star in the Catalog of PSPC
WGA Source of ROSAT~\citep{2000White}
\footnote{http://heasarc.gsfc.nasa.gov/W3Browse/rosat/wgacat.html}. 
The other sources at
$(\alpha, \delta)_{\rm J2000}=$ ($\timeform{18h27m36s}$,
$\timeform{-13D12m36s}$), ($\timeform{18h27m46s}$,
$\timeform{-13D13m22s}$), and ($\timeform{18h28m04s}$,
$\timeform{-13D11m25s}$) are not found in any X-ray source
catalogs.

\begin{figure}
\begin{center}
\FigureFile(80mm,50mm){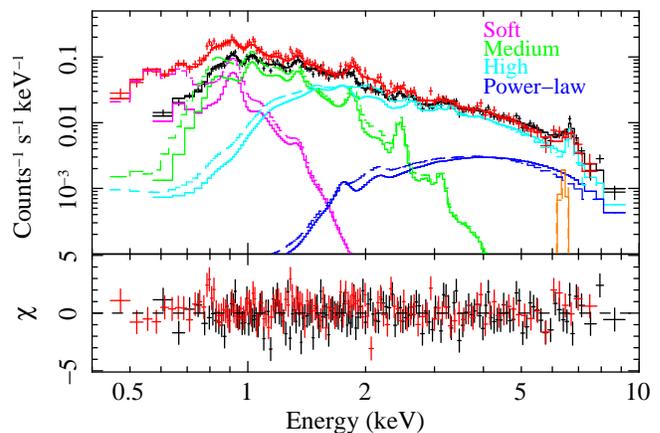}
\end{center}
\caption{
Spectra of the background region. The best-fit result of the model of thin 
thermal three-temperature plasma with an Fe\emissiontype{I} K$\alpha$ line plus the CXB 
is also shown.
Black and red data show the FI and BI spectra,
respectively. The vertical error bars of each data point are
the 1$\sigma$ error. The best-fit models for FIs and BI 
are shown with solid and broken lines, respectively. 
In this model, the soft, medium- and high-temperature 
components correspond to the GRXE.
The power-law component  corresponds to the CXB.  
The yellow line shows an Fe\emissiontype{I} K$\alpha$ line.

\label{fig:spectrumBG}}
\end{figure}

Excluding the four source regions of a \timeform{1.8'}-radius circle, 
we made the background spectrum from the data obtained by each XIS sensor 
in a circular with a radius of $\timeform{9'}$. 
The solid angle we extracted the spectra from is 212 arcmin$^2$. 
The Non-X-ray background (NXB) spectrum of each sensor was made from the NXB
database of the same region in the detector coordinate
(DETX/Y), sorted  by  the geomagnetic cut-off rigidity (COR) to become the same COR 
distribution of the background observation ({\tt xisnxbgen}).
This sorting process improves the reproducibility of the NXB spectrum up to
$<4\%$ accuracy (in detail, see \cite{Tawa2008}).  
We then subtracted the NXB spectrum from the background spectrum. 
Since the gains and responses of the
FIs are almost the same, we averaged the spectra of the three FIs. 
The background spectra are shown in figure~\ref{fig:spectrumBG}.

We fitted the spectrum (figure~\ref{fig:spectrumBG}) with a model consists of 
a thin thermal plasma component (APEC; \cite{2001Smith}) and the Cosmic X-ray Background (CXB) component.
The fitting were made for the FI and BI spectra simultaneously.
The detector responses (RMF) and telescope
responses (ARF) generated by {\tt xisrmfgen} and {\tt
xissimarfgen} \citep{Ishisaki2007} were averaged for the three FIs with {\tt addrmf} and {\tt addarf}. 

Since the CXB was not subtracted from the spectrum,
we added the CXB model spectrum in the fitting, assuming a power-law of photon index 1.4 
with the intensity of  $5.4 \times 10^{-15}$~erg~s$^{-1}$~cm$^{-2}$~arcmin$^{-2}$ (2--10~keV) \citep{Kushino2002}.   
The absorption for the CXB is the  Galactic H\emissiontype{I} column
density towards the background region of $\sim 1.1 \times
10^{22}$ $\mathrm{cm}^{-2}$~\citep{2005Kalberla}\footnote{We
used the H\emissiontype{I} calculator available at
http://heasarc.gsfc.nasa.gov/cgi-bin/Tools/w3nh/w3nh.pl.},
plus that of molecular hydrogen of $\sim 2 \times 10^{22}
~\mathrm{cm}^{-2}$~\citep{Dame2001}. 
Then the overall column density for the CXB ($N_{\rm H}^{\rm CXB}$) is 
$N\mathrm{_H^{total}} = N\mathrm{_{HI}} +2N\mathrm{_{H_2}} 
= 5 \times 10^{22}~\mathrm{cm}^{-2}$.  
For a thin thermal plasma, the column density, temperature, metal abundances,
and normalization were free parameters.
The cross section of the photoelectric absorption and the solar abundance of each
element were obtained from \citet{Morrison1983} and \citet{And89}.

This model, however, is rejected with $\chi^2$/d.o.f.=1117.1/280.
A model of two-temperature plasma plus the CXB was also rejected ($\chi^2$/d.o.f.=609.9/277) 
with a large residual at 0.56~keV, which is attributable to the O\emissiontype{VII} K$\alpha$ line. 
We then fitted the spectra with three-temperature plasma plus the CXB model, where the
abundances in each plasma were assumed to be the same.  The fit was improved to be 
$\chi^2$/d.o.f.=371.1/274, but still large residuals were seen in 
Ne\emissiontype{X}, S\emissiontype{XV} and 
Ar\emissiontype{XVII} K$\alpha$ lines. Then we relaxed the abundances of 
Ne, S and Ar to be free, and the fit was acceptable with $\chi^2$/d.o.f.=310.5/271.

The main component of this spectrum is thought to be the GRXE. 
An Fe\emissiontype{I} K$\alpha$ line is  reported in the GRXE \citep{Ebisawa2008},
In fact, if we limited the fitting energy range to be 5.5--7.4 keV,
the $\chi^2$/d.o.f. were 35.9/17 and 27.1/16, with and without the Fe\emissiontype{I} K$\alpha$ line, respectively.
We thus added the Fe\emissiontype{I} K$\alpha$ line in the final model.
The best-fit results are shown in table~\ref{tab:BGfitting} and figure~\ref{fig:spectrumBG}.

\begin{table}
\caption{Result of the spectrum fitting of the background region with the model of thin 
thermal three-temperature plasma with an Fe\emissiontype{I} K$\alpha$ line plus the CXB.\label{tab:BGfitting}}
\begin{center}
\begin{tabular}{lccc}
 \hline \hline
& Soft & Medium & High\\
\hline
$N_{\rm H}$ ($10^{22}$~cm$^{-2}$) & $0.61^{+0.21}_{-0.11}$ & $0.84^{+0.09}_{-0.08}$ &  $1.19^{+0.23}_{-0.25}$ \\
$kT$ (keV) & $0.11^{+0.03}_{-0.03}$ &  $0.50^{+0.04}_{-0.03}$ & $4.76^{+0.65}_{-0.55}$ \\
$f_{\rm 0.8-10}$\footnotemark[$*$] ($10^{-13}$) & 1.8 (10) & 8.4 (47) &   30 (44)  \\

&\multicolumn{3}{c}{Abundance (solar)}\\

Ne & \multicolumn{3}{c}{$0.57^{+0.19}_{-0.12}$} \\
S  &  \multicolumn{3}{c}{$1.27^{+0.56}_{-0.33}$} \\
Ar &  \multicolumn{3}{c}{$2.10^{+1.29}_{-1.03}$} \\
others &  \multicolumn{3}{c}{$0.33^{+0.09}_{-0.07}$}\\ 
\hline
$f_{\rm 6.4}$\footnotemark[$\dag$] ($10^{-6}$) &\multicolumn{3}{c}{$3.5\pm2.0$}\\ 
\hline
$\chi^2$/d.o.f. & \multicolumn{3}{c}{302.8/270}\\ \hline
\end{tabular}
\end{center}
$*$ Observed flux in the 0.8--10 keV band in the unit of erg~s$^{-1}$~cm$^{-2}$. Values in
parentheses are the absorption corrected values.\\
$\dag$ Observed photon flux of 6.4~keV line in the unit of photons~s$^{-1}$~cm$^{-2}$.
\end{table}

The temperatures and column densities of the medium- and
high-temperature components are similar to those of the
GRXE~\citep{Ebisawa2005}.  
Although the soft component has the same temperature as the local hot bubble, 
the absorption column is larger than the local hot bubble,
and similar to  that of the medium component of 
$\sim10^{22}$~cm$^{-2}$. Thus the soft component may have a comparable distance 
of the low-temperature component of the GRXE, 
and hence may comprise additional component of the GRXE.
To confirm the same distance, we tried a model with a common 
$N_\mathrm{H}$ for the soft and medium-temperature components.
This model was acceptable with $\chi^2$/d.o.f.= 310.3/271. 
The common $N_\mathrm{H}$ is 0.81$\pm 0.06\times10^{22}~$cm$^{-2}$.

In the 0.8--10 keV band, the contribution of the low-temperature component is only 4\%, 
we hence used  the model GRXE  spectrum as being the medium and high components, at least
above 0.8~keV. 

\subsection{Source Region}
\subsubsection{Image\label{ch:image}}

\begin{figure} 
\begin{center}
(a)\FigureFile(80mm,60mm){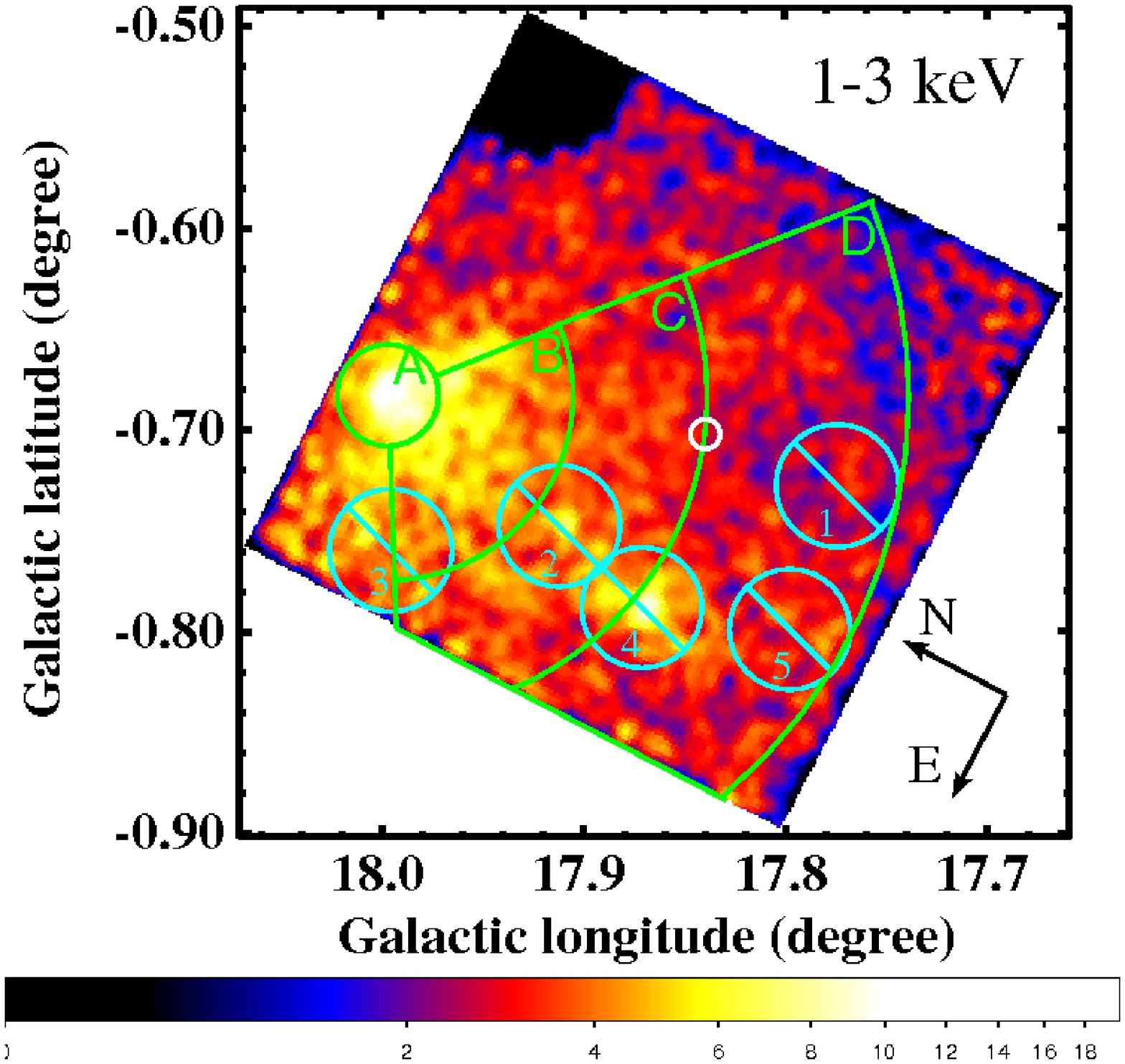}\\
(b)\FigureFile(80mm,60mm){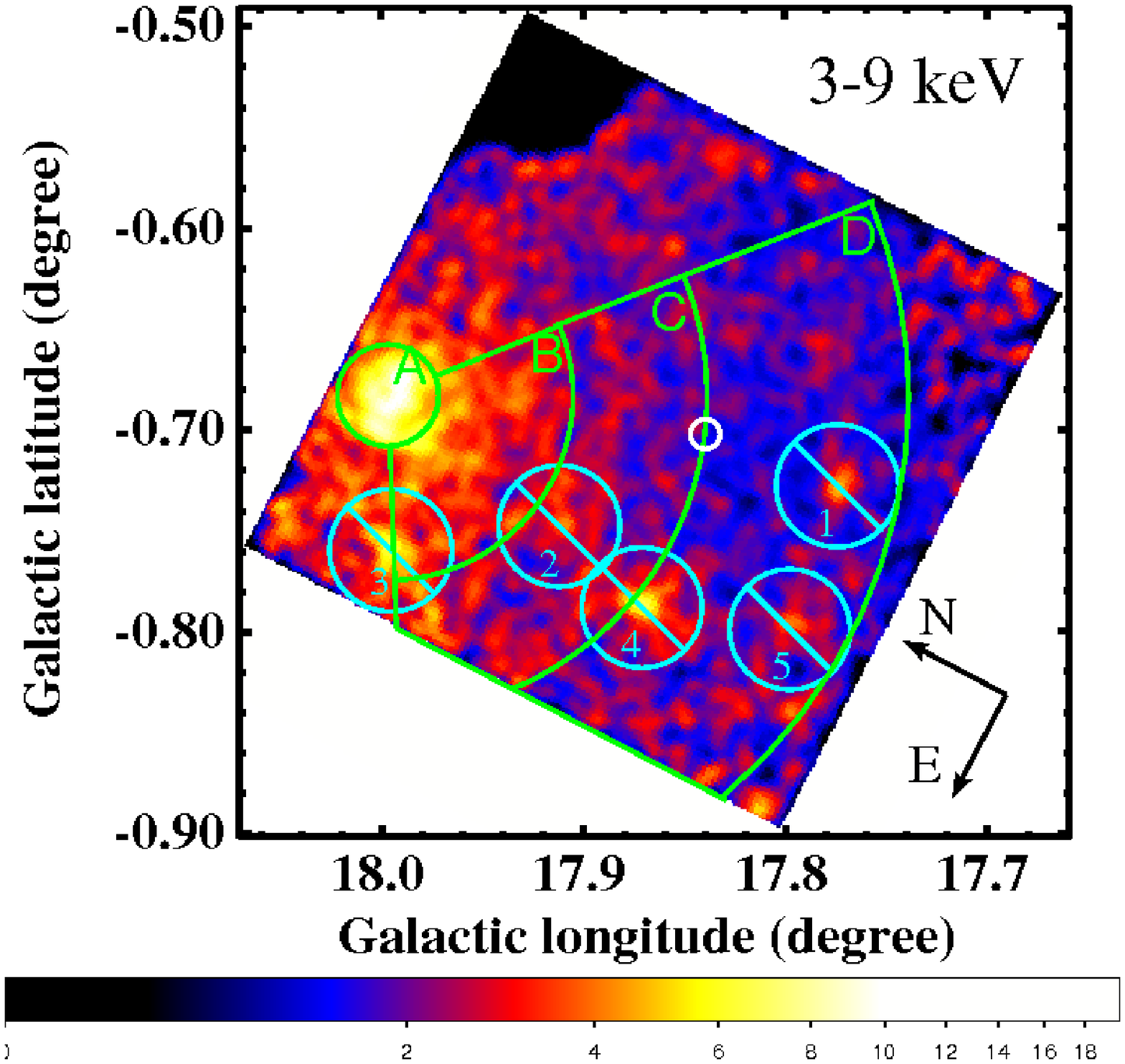}
\end{center}
\caption{
XIS images of the source region in the 1.0--3.0~keV (a) and
3.0--9.0~keV (b) bands in the Galactic coordinate. The NXB
was subtracted and the vignetting was corrected. These
images were smoothed using a Gauss function with $\sigma =
\timeform{0.2'}$.  The scale is in units of photons. A
corner illuminated by the calibration source was removed.  
The peak position of the VHE ${\gamma}$-ray emission is shown
with the white circle. The source regions of spectra
(regions A, B, C and D) are shown with the green lines. 
The cyan circles with diagonal lines (1--5) are the point-like source
regions excluded from the analyses of the spectra and the radial
profile. The sources 1--4 are cataloged by XMM-Newton.
\label{fig:image}}
\end{figure}

The image of the source region was made with the vignetting correction ({\tt xissim}; Ishisaki~et al.~2007) 
after the NXB  subtraction. Figure~\ref{fig:image} shows the 3-FI images of the source region in 
the 1--3~keV and 3--9~keV bands.  

Among the five bright point-like sources in figure~\ref{fig:image}, four are identified in the XMM-Newton
serendipitous source catalogue (2XMM)\footnote{
http://xmmssc-www.star.le.ac.uk/Catalogue/2XMM/}:
2XMM~J182557.9--134755, 2XMM~J182620.9--134426,
2XMM~J182617.1--134111 and 2XMM~J182629.5--133648. 
Referring the accurate  position of these XMM-Newton 
sources ($\leq$\timeform{1"})\footnote{We referred to XMM-Newton Users' Handbook, available at
\linebreak http://xmm.vilspa.esa.es/external/xmm\_user\_support/documentation/uhb\_2.5/index.html.},
we corrected the coordinates of XIS images by shifting 
$\timeform{24"}$ to match the source positions with each other. 
After this correction, the position of
the brightest peak in figure~\ref{fig:image} coincides with
PSR~J1826$-$1334. 
The coordinate of the point source not cataloged by XMM-Newton (source 5 in figure \ref{fig:image}) 
is $(\alpha, \delta)_{\rm J2000}=$ ($\timeform{18h26m12s}$, $\timeform{-13D48m33s}$) after the correction.

Although XMM-Newton~\citep{Gaensler2003PSRB1823} and
Chandra~\citep{Pavlov2008} resolved the extended
emission around PSR~J1826$-$1334 into the core and
diffuse components,
the XIS could not resolve due to the limited spatial resolution. 
We, instead,  found  that the diffuse emission is largely extended beyond \timeform{5'}, 
which is the radius previously reported.

In order to study the spatial distribution of the diffuse component, 
we made the radial profile of the surface brightness in the 1.0--9.0~keV band. 
The origin of the profile is at the pulsar point, 
and the data area of the profile is shown in figure~\ref{fig:obspos}.  
Fluxes of the five point sources in the circles of $\timeform{1.8\prime}$ radius
were removed (figure~\ref{fig:image}).

\begin{figure}
\begin{center}
\FigureFile(80mm,50mm){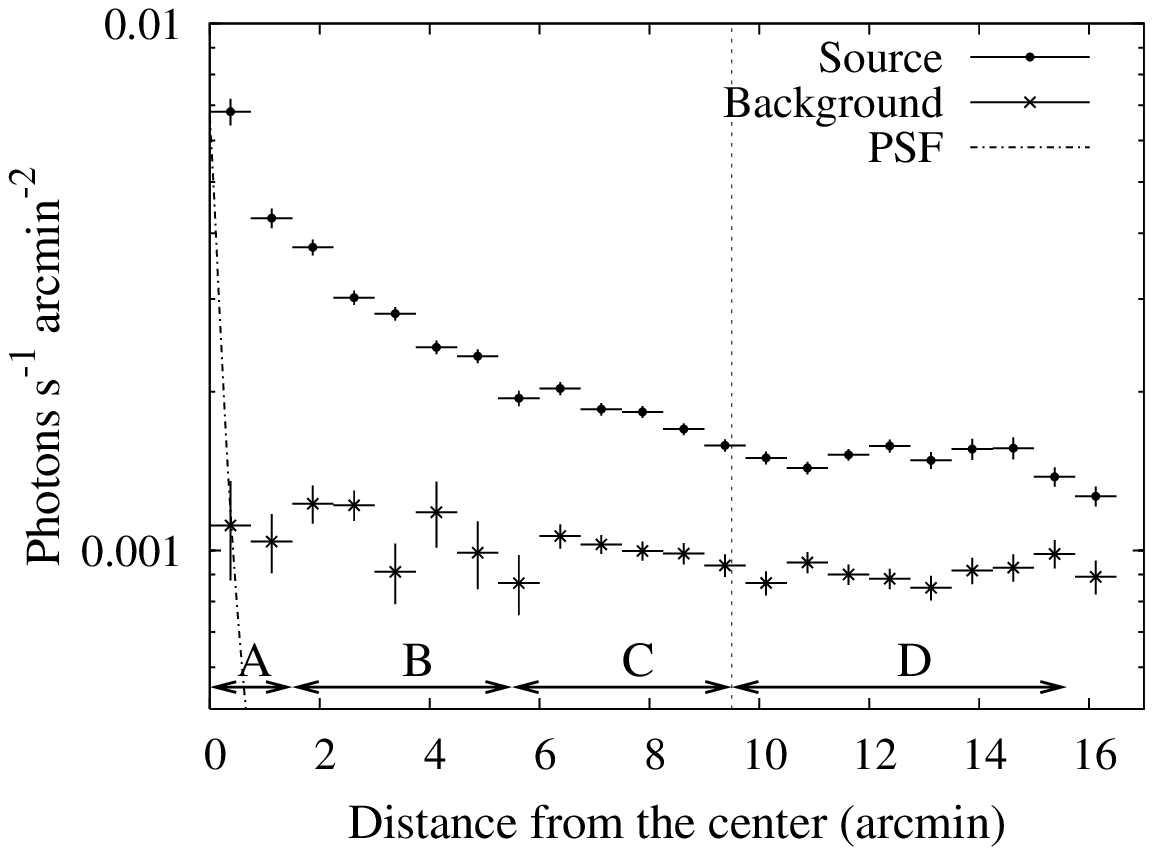}
\end{center}
\caption{
Radial profile of the surface brightness per 1~FI CCD in the 1.0--9.0 keV
band. The NXB was subtracted and the vignetting was corrected. 
The vertical error bar of each
data point is the 1$\sigma$ error. The position of the peak of
VHE ${\gamma}$-ray emission is shown by a vertical broken
line.\label{fig:radialprofile}}
\end{figure}

For comparison, we also made the radial profile in the same detector area  of the background region with the same process as the source profile (see figure~\ref{fig:obspos}).
These two profiles are shown  in figure~\ref{fig:radialprofile}. The surface brightness of the source region is clearly larger than that of the background region.

\subsubsection{Spectrum\label{ch:spectroscopy}}

We made X-ray spectra from the four regions designated as A, B, C, and D in figure 3.  
Like the radial profile, we removed the data of the five point sources.
The region A is a circle of $1^{\prime}.5$-radius centered on PSR~J1826$-$1334. 
Region A includes the core region, while region B corresponds to the diffuse region, both are reported by
\citet{Gaensler2003PSRB1823}.  Regions C and D
are new diffuse X-ray area discovered by this Suzaku
observation. 

\begin{figure*}
\begin{center}
\FigureFile(80mm,50mm){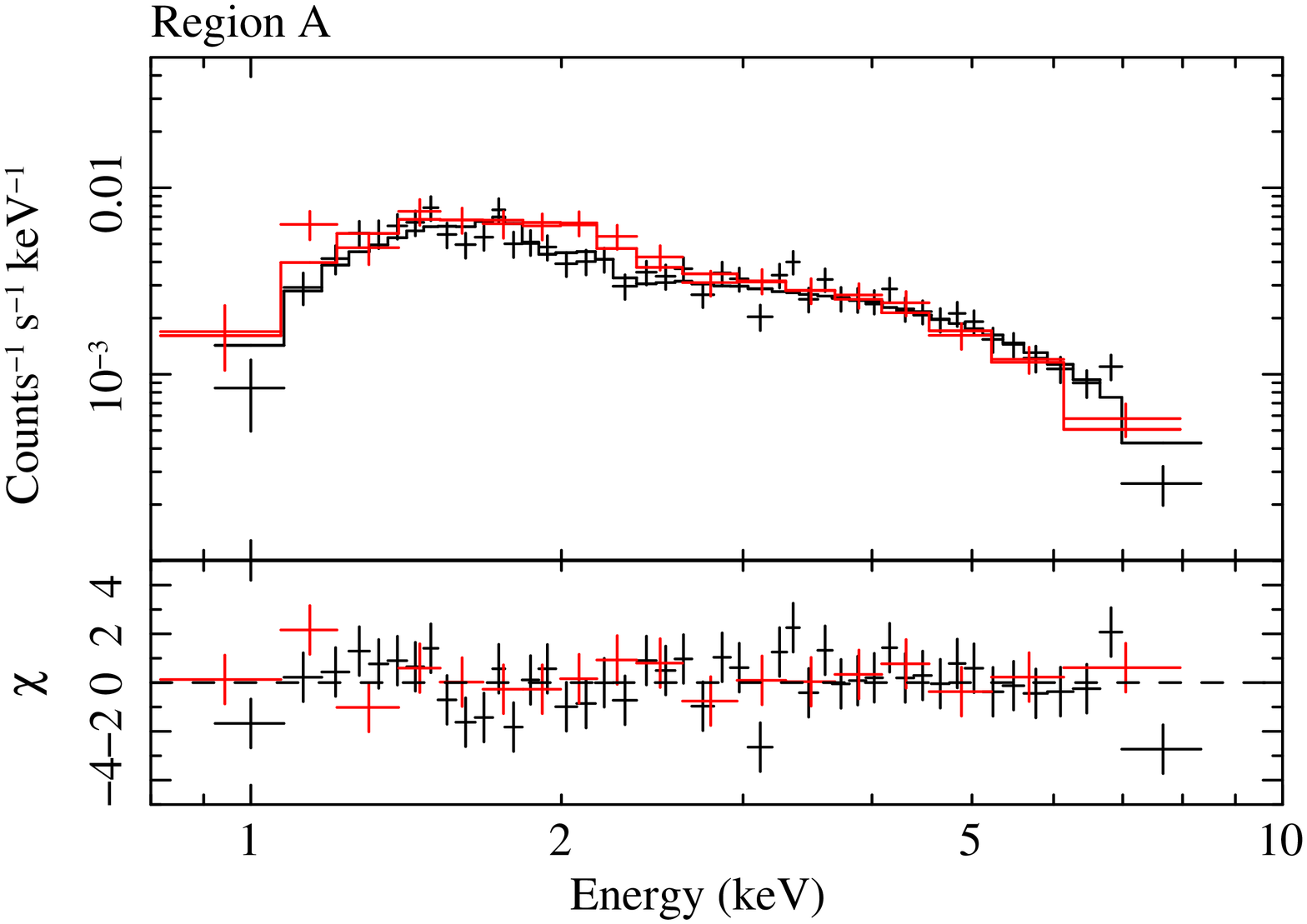}
~~~~
\FigureFile(80mm,50mm){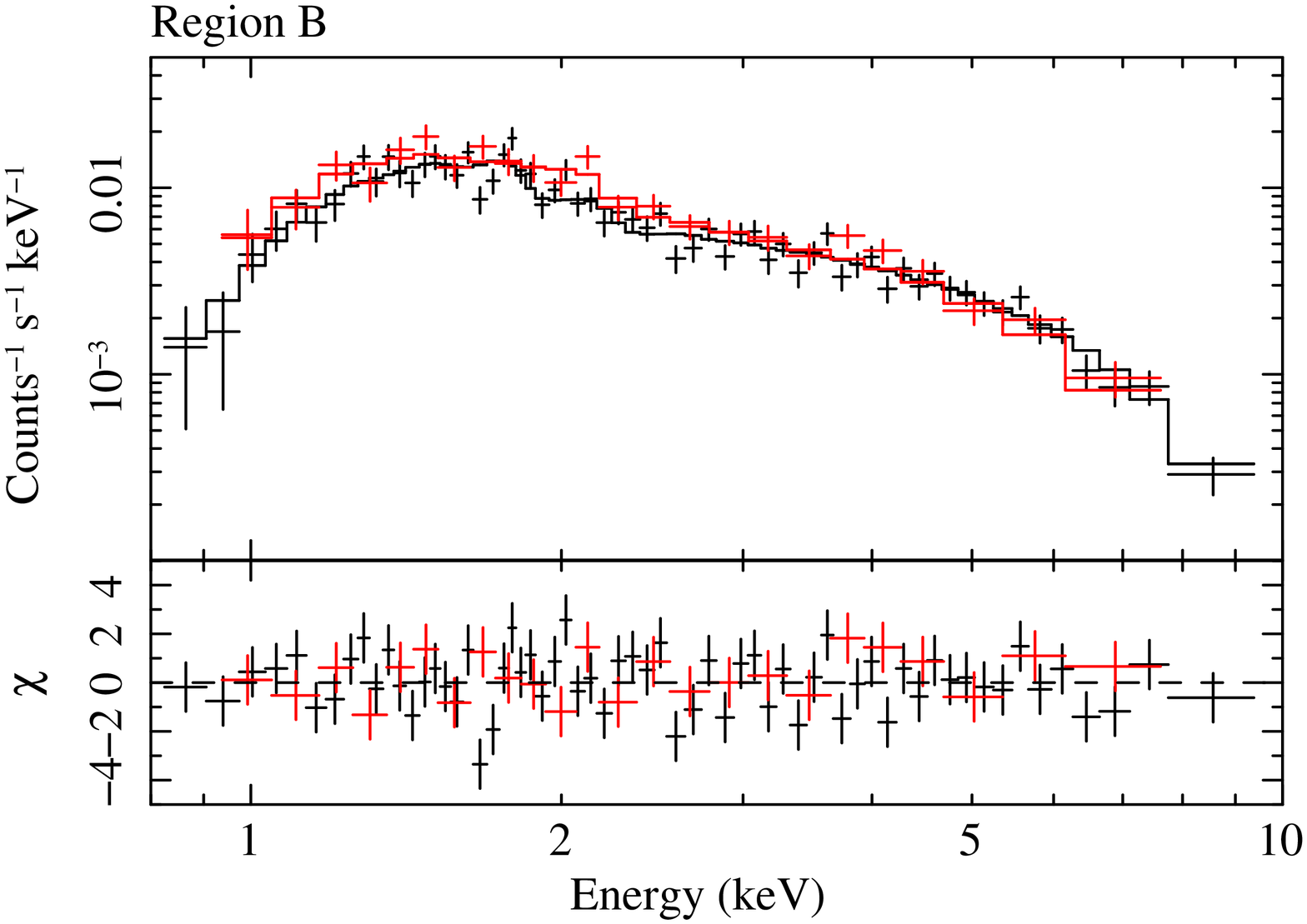}
\FigureFile(80mm,50mm){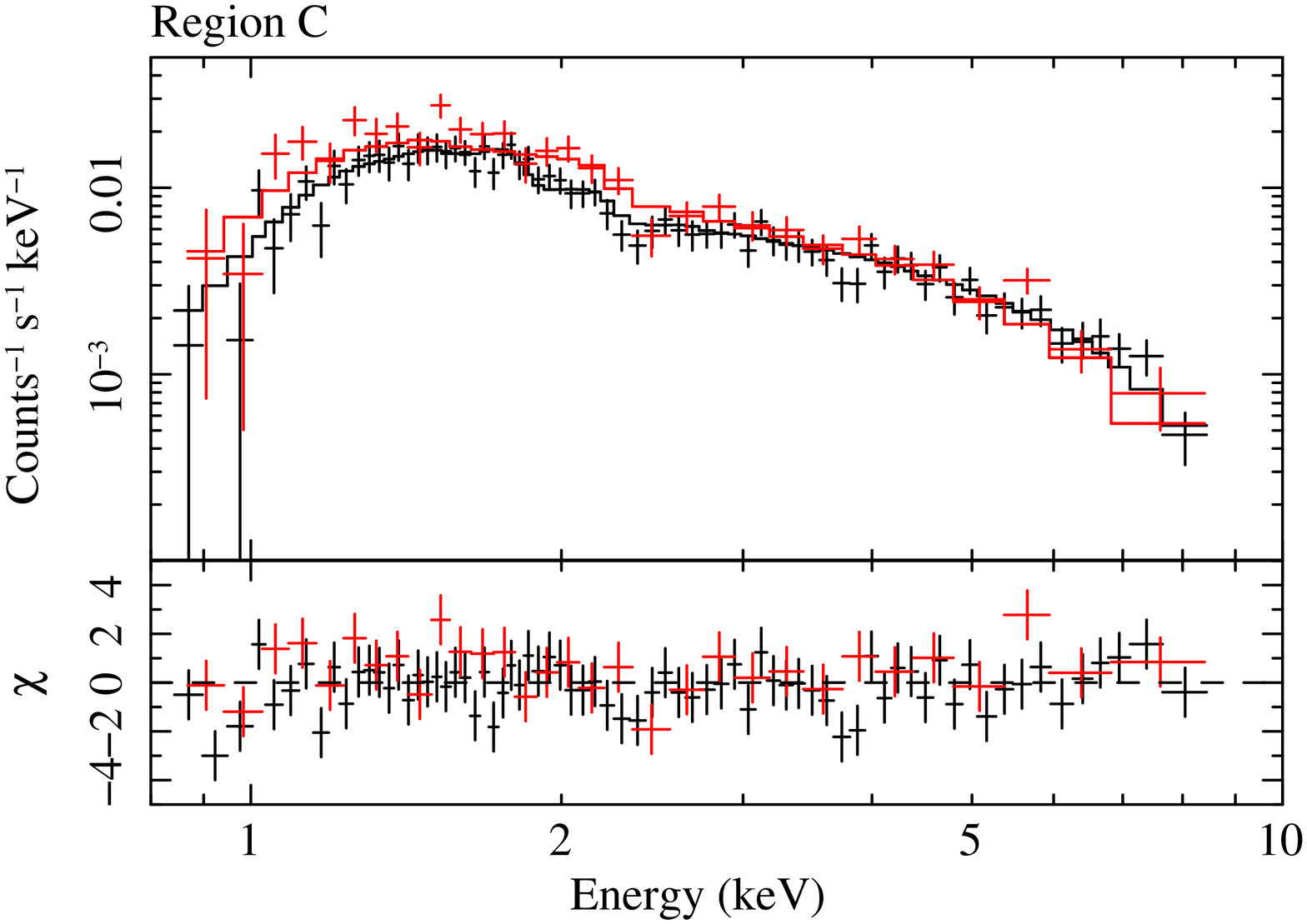}
~~~~
\FigureFile(80mm,50mm){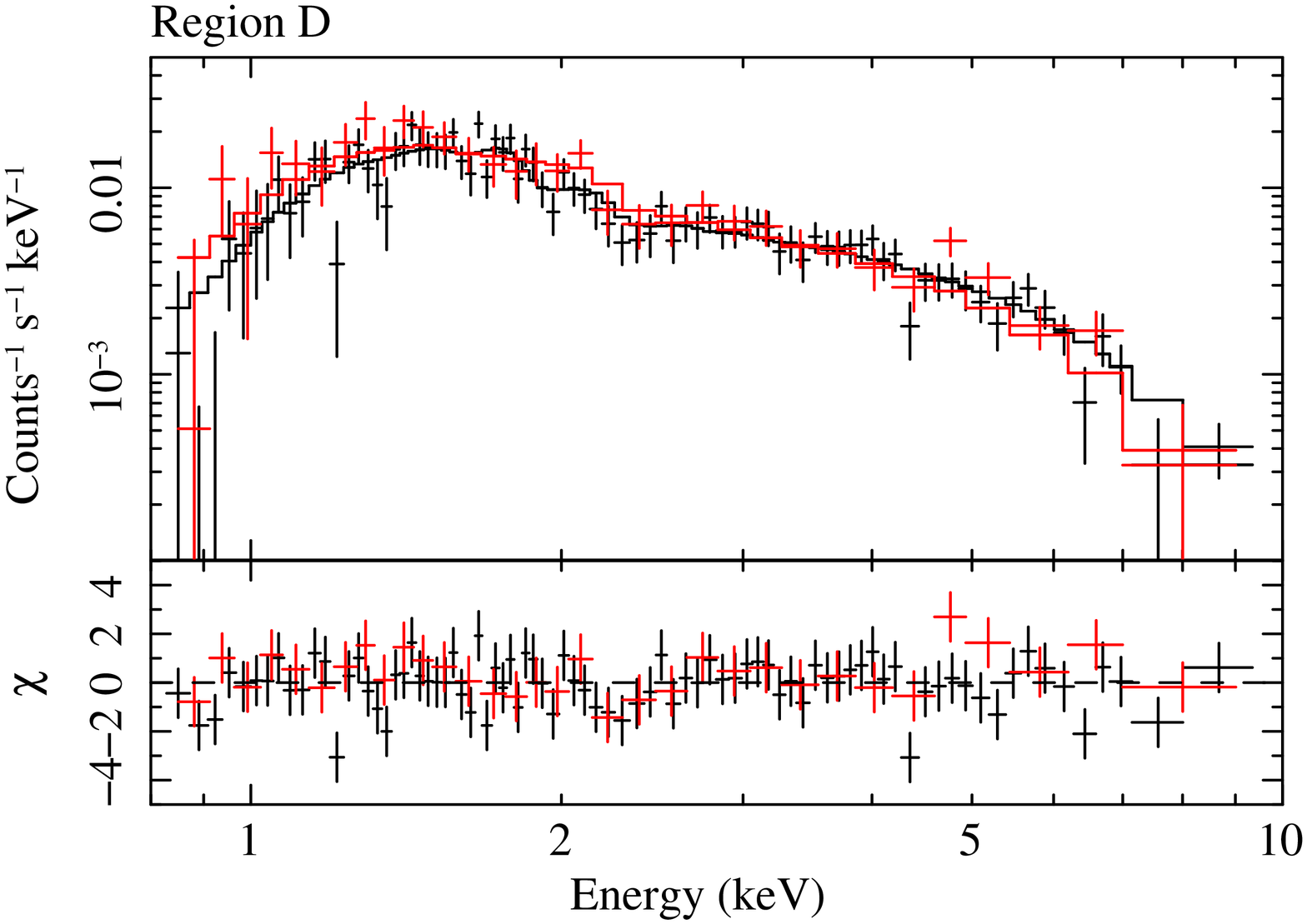}
\end{center}
\caption{
Spectra of the regions A, B, C and D with the best-fit
models. Black and red data show the spectra of the FI and
BI, respectively. The vertical error bars of each data point
are the 1$\sigma$ error.
\label{fig:spectra}}
\end{figure*}

The spectra were made with the same process as the background region,
and subtracted the background spectrum shown in figure~\ref{fig:spectrumBG}. 
In the background subtraction, we
took into account the difference of the vignetting between
the regions A--D and the background region.  The
background-subtracted spectra are shown in figure~\ref{fig:spectra}.

Since the spectra have featureless with no  emission line, we 
fitted the spectra with an absorbed power-law model. Free parameters are the normalization, photon index ($\Gamma$), and $N\mathrm{_H}$.  The fits are acceptable with the best-fit results shown in figure~\ref{fig:spectra} and table~\ref{tab:fittings}.

\begin{table*}
\begin{center}
\caption{Best-fit results of the regions A, B, C and D.\label{tab:fittings}}
\begin{tabular}{ccccccc} \hline \hline
Region\footnotemark[$*$]	& $\Omega$\footnotemark[$\dagger$] &$N_{\mathrm{H}}$ ($10^{22}~\mathrm{cm}^{-2}$ )	&  $\Gamma$\footnotemark[$\ddagger$] & Norm.\footnotemark[$\S$] ($10^{-4}$) & $f_{\mathrm{0.8-10}}$\footnotemark[$\l$] ($10^{-12}$)  & ${\chi}^2$/d.o.f. \\ \hline	
A & 7.07 & $1.11_{-0.15}^{+0.14} $ & $1.69_{-0.07}^{+0.09} $ & $3.0_{-0.5}^{+0.5}$ & 1.3 (1.7) & 65.1/61 \\
B & 21.4 & $0.98_{-0.12}^{+0.11} $ & $1.97_{-0.09}^{+0.09} $ & $4.2_{-0.6}^{+0.6}$  & 1.2  (1.7) &   104.6/84\\
C & 39.8 & $0.91_{-0.11}^{+0.12} $ & $2.01_{-0.10}^{+0.10} $ & $4.4_{-0.7}^{+0.7}$  & 1.2 (1.7) &  105.0/100 \\
D & 76.3 &  $0.86_{-0.15}^{+0.14} $ & $1.94_{-0.08}^{+0.12} $ &  $4.5_{-0.8}^{+0.9}$  & 1.4 (1.9) & 111.1/113 \\ \hline
\end{tabular}
\end{center}			    
\footnotemark[$*$] The regions are shown in figure \ref{fig:image}.\\
\footnotemark[$\dagger$] Solid angle of the region ; arcmin$^{2}$.\\
\footnotemark[$\ddagger$] Photon index of the power-law model.\\
\footnotemark[$\S$] Normalization of the power-law model ; photons keV$^{-1}$ s$^{-1}$ cm$^{-2}$ at 1 keV.\\
\footnotemark[$\l$] Observed flux in the 0.8--10 keV band in the unit of erg~s$^{-1}$~cm$^{-2}$. Values in
parentheses are the absorption corrected values.\\
\end{table*}

The power-law index of the region A is $\Gamma=1.7$, which  is harder than the other 3 regions.  
The regions B, C and D show almost the same photon indices of  $\Gamma=2.0$  and the same  interstellar absorptions 
of $N_{\mathrm{H}} \sim 1\times 10^{22}~\mathrm{cm}^{-2}$.
We therefore added these 3 spectra, and examined whether the spectrum contains thermal component.
Most probable thermal plasma, if any, in the extended region, would be the shock heated interstellar gas. 
We therefore fitted the spectrum with a thin thermal model of solar abundance.
This model is rejected with $\chi^2$/d.o.f.= 547.0/301. 
Then we relaxed the abundance to be free, and found that the model becomes acceptable with
$\chi^2$/d.o.f.= 341.8/300. However we need an unrealistically low abundance of less than 0.15 solar.  
We hence can say  that the thin thermal plasma of solar abundance contributes, at most, only 15\% 
in the 0.8--10 keV flux of the largely extended emission. 
The best-fit parameters of these fitting are shown in table \ref{tab:testmodel}.  

\begin{table}
\begin{center}
\caption{Best fit parameters of  the fittings of the combined spectra of the region B--D with different models.\label{tab:testmodel}}
\begin{tabular}{cccc} \hline \hline
Model &  Power law & APEC & APEC \\
\hline
$\Gamma$/$kT$(keV) &  $1.98_{-0.06}^{+0.05}$ & $13_{-2}^{+1} $   &  $6.4_{-0.7}^{+0.7}$ \\  
Abundance (Solar) &  $-$ & 1 (fixed)	&  $<$0.15 \\
$N_{\mathrm{H}}$ ($10^{22}~\mathrm{cm}^{-2}$ ) &  $0.93_{-0.09}^{+0.06}$ &  $0.47_{-0.04}^{+0.03}$ &  $0.70_{-0.03}^{+0.05}$ \\
\hline
${\chi}^2/$d.o.f. & 326.0/301 & 547.0/301  &341.8/300\\
\hline	
\end{tabular}
\end{center}
\end{table}

\section{Discussion}
\subsection{Radial Profile}

In figure~\ref{fig:radialprofile}, we see largely extended emission 
up to \timeform{15'} for the first time. The radial profile becomes flat beyond \timeform{10'}. 
The flux in this flat region is 50\% larger than the mean flux of the background region.  
One may argue that the difference could be due to the fluctuations of the NXB,  CXB and GRXE.  
We therefore discuss  whether any  systematic errors of the background subtraction (NXB, CXB and GRXE) may mimic such excess. 

The reproductively of the NXB is extensively examined by \citet{Tawa2008}. Using the same method, 
the NXB reproductively in the 1.0--9.0 keV band at the 99\% confidence level is less than
$\mathrm{2.2 \times10^{-5}~counts~s^{-1}~arcmin^{-2}}$.
This uncertainty is very  small compared with the excess level of the radial profile of the source region (figure~\ref{fig:radialprofile}).

The fluctuation of the CXB in the XIS FOV is estimated to be $\sim$ 36\% at the 99\% confidence 
level with the same way as
\citet{Tawa2008}.  Since the CXB flux is only 10\% of the background 
spectrum, the overall uncertainty is less than 4\%.

The largest uncertainty would come from the subtraction of the GRXE. 
\citet{Yam93} and \citet{Kan97} reported that the GRXE has a large scale structure
depending on the galactic longitude ($l$) and latitude ($b$). The $l$-dependency is rather null
in this region hence can be neglected in our case. The $b$-dependency is expressed by the scale height
of the medium and hard components given by 2 and 0.7 degrees, respectively \citep{Yam93, Kan97}.
Since the flux of the medium and hard components in our region is nearly the same in the 1.0-9.0 keV band, 
we can estimate that GRXE flux at the source position is about 20\% larger than that in the background position. 

The most unknown factor is a possible fluctuation of the GRXE from position to position, which 
may cause under/over subtraction of the GRXE. 
If there is under/over subtraction of the GRXE, the source spectra should have emission/absorption line 
structures at the Si, S and Fe K-shell transition energies, because the GRXE has strong lines
of these elements (see figure \ref{fig:spectrumBG}).
No such line/absorption structures are found in the spectra of the regions B, C, and D given in figure 5, which
supports that the GRXE is properly subtracted.
We further checked possible under/over subtraction of the GRXE by fitting the source spectrum (combined
B+C+D) with a model of absorbed power-law + $\Delta\times$ GRXE, where the spectral parameters of the GRXE, 
other than normalization
($\Delta$) were fixed to the best-fit values given in table~\ref{tab:BGfitting}. 
As a result, we can set the 90\% upper-limit of the possible contamination of the GRXE to be 1.5\% 
of the excess flux.
Accordingly, the largely extended power-law component from the source region is
real, not an artifact of improper GRXE subtraction.

\subsection{Comparison with earlier X-ray observation}
In the XMM-Newton observation, \citet{Gaensler2003PSRB1823} reported that the photon indices 
of the core and diffuse components are $\sim$ 1.6 and $\sim$ 2, respectively. Since the region A includes
both the components, our result of $\Gamma \sim 1.7$ is reasonable. 

From the region B, we determined more accurate  photon index and interstellar absorption than 
the diffuse component of XMM-Newton.
Moreover, we detected the X-ray emission farther out of the
previously reported region. The peak position of the extended
emission coincides with the pulsar PSR~J1826$-$1334, and the
surface brightness decreases with the distance from the
pulsar. The major fraction of the extended emission cannot be
a thin thermal plasma of the SNR, but is non-thermal origin. 
The photon indices are smaller
than the canonical value of synchrotron emissions found in
non-thermal SNRs (2.5--3.0; Bamba~et al.~2005).  Furthermore,
the morphology of the extended emission is different from
limb-brightened structures such as
SN1006~\citep{Koyama1995}.  Thus the extended emission cannot be
explained as non-thermal emission from shell-like SNR. 
The photon indices are close to 
typical values of PWNe~\citep{Gaensler2006}.
Therefore, we conclude that the extended emission is synchrotron radiation
from a PWN, as suggested first by \citet{Gaensler2003PSRB1823}.

Our results indicate that the PWN extends over $\timeform{15'}$,
which corresponds to 17$d_\mathrm{4kpc}$~pc, where
$d_\mathrm{4kpc}$ is the distance of PSR~J1826$-$1334
normalized by 4~kpc. The interstellar absorption
of $N_{\mathrm{H}} \sim 10^{22}
\mathrm{cm}^{-2}$ is consistent with the
distance of $\sim 4$~kpc obtained by the radio
measurements~\citep{Taylor1993,Cordes2002}. The extent of
$\sim 17d_{\rm 4kpc}$~kpc is larger than typical
X-ray PWNe; for example, one of the largest PWNe is 3C58,
where the X-ray nebula extends up to $\sim 6$~pc from the
pulsar~\citep{Slane2004}.

The total X-ray luminosity ($L_{\rm X}$) from the regions
A--D are $8.6 d_{\rm 4kpc}^2 \times 10^{33}$~erg~s$^{-1}$ in
the 2--10 keV band. The ratio of $L_{\rm X}$ to the
spin-down luminosity is not unusual compared with other
PWNe~\citep{Che04,Li07}.

\subsection{Comparison with $\gamma$-ray observation}

\begin{figure} 
\begin{center}
\FigureFile(80mm,60mm){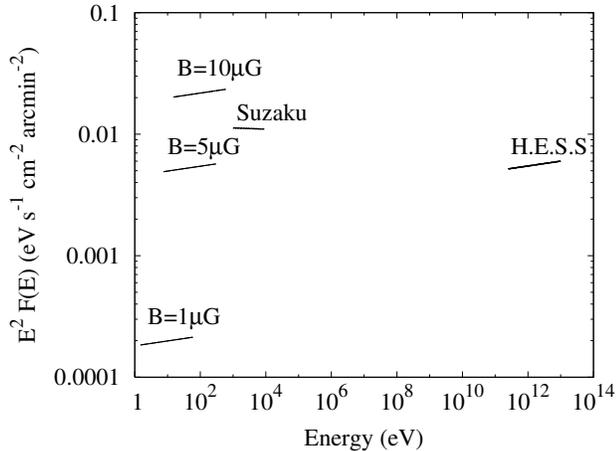}
\end{center}
\caption{
Spectral energy distribution of the extended component from
the X-ray to TeV $\gamma$-ray bands. The intensity of the
region C in table~\ref{tab:fittings} and that of the Radius
\timeform{0.15D} region in table~2 of
\citet{Aharonian2006HESSJ1825} are used for X-ray and VHE $\gamma$-ray.
The synchrotron radiation 
from the electrons responsible for VHE $\gamma$-ray (0.25$-$10~TeV) is 
plotted toward the left for three different values of the magnetic field. 
\label{fig:SED}}
\end{figure}

Since the region C overlaps the bright region of 
the VHE $\gamma$-rays, we can make  a reliable 
spectral energy distribution (SED) between
the X-rays and the VHE $\gamma$-rays for the first time.
In figure~\ref{fig:SED}, we show the X-ray SED of the region
C and the $\gamma$-ray SED of the brightest region 
(Radius \timeform{0.15D} in table~2
of \cite{Aharonian2006HESSJ1825}). 

If the origin of the VHE $\gamma$-rays is the inverse
Compton scattering of the CMB photons by high-energy
electrons, these electrons should emit synchrotron radiation.
We also plot the estimated synchrotron intensity in the
magnetic fields of $B= 10,~5,~1~\mu$G.  In the case of $B
\sim 7~\mu$G, the estimated synchrotron intensity smoothly connects
to the X-ray intensity.  Thus both the X-rays and VHE
$\gamma$-rays can be explained by high-energy electrons of a
single population in a magnetic field of $B \sim 7~\mu$G, 
approximately the same as the core region of $B \sim 10~\mu$G 
~\citep{Gaensler2003PSRB1823}. 

According to \citet{deJager2008}, required electron energy
$E_\mathrm{e}$ in a magnetic field $B$ to radiate
synchrotron photons of mean energy $E_\mathrm{syn}$ is given
by $E_\mathrm{e}$ $=$ $120~\mathrm{TeV}~(B/7\mu
\mathrm{G})^{-1/2} (E_\mathrm{syn} / \mathrm{2keV})^{1/2}$.
On the other hand, the mean electron energy $E_\mathrm{e}$
required to inverse Compton scatter the CMB photons to
energies $E_\mathrm{IC}$ is typically
$E_\mathrm{e}=18~\mathrm{TeV}~(E_\mathrm{IC}
/\mathrm{1TeV})^{1/2}$ \citep{deJager2008}. Thus electrons
responsible for X-rays have larger energies than those for
VHE $\gamma$-rays.
 
The peak position of the VHE $\gamma$-rays does not coincide
with the pulsar~\citep{Aharonian2006HESSJ1825}, while the
peak position in X-rays are consistent with the pulsar.
There is no indication of the $\gamma$-ray peak in the XIS
images (figures~\ref{fig:image} and
\ref{fig:radialprofile}). The offset is a mystery, and the reason
is not yet clarified. However, as \citet{Pavlov2008}
mentioned, the local excess of the seed photons for the
inverse Compton scattering created by the near hidden star
might cause the offset. In this case, a magnetic field
stronger than $\sim 10~\mu$G might be required.

\citet{Har99} reports an EGRET source 3~EG J1826--1302 
in the north of PSR~J1826--1334, and the relation between 
them has been discussed  by some authors (e.g. Zhang, Chen \& 
Fang~2008). The EGRET source is , however, out of the FOV of the XIS 
in our observation. Thus direct comparison between the intensities of the EGRET 
source and the X-ray we found is impossible, and the relation 
between them is not clarified.

\subsection{Widely extended PWN}

Except for the neighborhood of the pulsar, the photon index
is constant at $\Gamma=2.0$ from the pulsar to $\timeform{15'}$. 
The smooth decay of the intensity with the distance
from the pulsar shown in figure~\ref{fig:radialprofile}
suggests that the accelerator is the pulsar itself.  This
means the high-energy electrons should be transported over
$17d_\mathrm{4kpc}$~pc within its synchrotron life time.
According to \citet{deJager2008}, the synchrotron life time
of an electron emitting photons of mean energy
$E_\mathrm{syn}$ in an isotropic magnetic field of the
strength $B$ is $\tau_\mathrm{syn}$ $=$ $1.9~{\rm
kyr}~(B/7\mu \mathrm{G})^{-3/2} (E_\mathrm{syn} /
\mathrm{2keV})^{-1/2}$.  Thus, the velocity for transporting
the high-energy electrons should be more than $17
d_\mathrm{4kpc}$~pc$/ \tau_\mathrm{syn}$ $\sim$ $8.8 \times 10^3~{\rm
km~s}^{-1} d_\mathrm{4kpc} (B/7\mu \mathrm{G})^{3/2}
(E_\mathrm{syn} / \mathrm{2keV})^{1/2}$.

A simple transport mechanism is diffusion by a magnetic field or convection. 
In the case of diffusion, our results indicate that the diffusion coefficient is
$D=R^2/(2\tau) = 2.3 \times 10^{28}
\mathrm{~cm^2~s^{-1}} (R/\mathrm{17pc})^2
(\tau/1.9\mathrm{kyr})^{-1}$, where $R$ is the size of the
extended emission, and $\tau$ is the transport time for
which we assume the synchrotron life time.

We can express the mean free path of the electron in a magnetic field
$B$ as $f r_\mathrm{L}$, where $r_\mathrm{L} = E_e/(eB)$ is
the gyro radius of an electron with energy $E_e$ and charge
$e$, and the parameter $f$ characterizes the efficiency of
diffusion. Then the diffusion coefficient can also be expressed as
$D$ $\sim$ $f r_\mathrm{L} c /3$ $=$ 
$2.3\times 10^{28} \mathrm{~cm^2~s^{-1}}~(B/7\mu
\mathrm{G})^{-1}~(E_e/120\mathrm{TeV}) (f/40)$.
Thus, our results suggest $f \sim 40$. According to
\citet{deJager2008}, $f$ should be $\le 1$ in perpendicular
diffusion. Therefore, our results suggest that the transport
mechanism is not the perpendicular diffusion, but it might
be diffusion parallel to the magnetic field or convection.

Collisions between a PWN and a reverse shock from the
surrounding SNR were proposed as a scenario to explain the
morphology of one-sided PWNe \citep{Pavlov2008}, such as Vela
pulsar~\citep{Blondin2001, Gaensler2003PSRB1823}.
Whether or not this scenario can explain the large extent of
$\sim$17$d_{\rm 4kc}$~pc and the smooth radial profile shown
in figure~\ref{fig:radialprofile} is an open problem.  

\section{Summary}
\begin{itemize}

\item{We observed the peak position of HESS~J1825$-$137 with the
Suzaku XIS, and found diffuse X-ray emission extended at
least up to $\timeform{15'}$ corresponding to 17~pc. } 

\item{The spatially-resolved X-ray spectra have no emission line, and can be fitted with an absorbed power-law
model. Therefore the X-rays are likely to be synchrotron
emission from a pulsar wind nebula.  Except for the
neighborhood of the pulsar, the photon indices are spatially
constant at $\Gamma=2.0$. } 

\item{The SED of the X-ray and the VHE $\gamma$-rays is made from the same region,
which can be explained by high-energy electrons of a
single population in a magnetic field of $B
\sim 7~\mu$G. }

\item {All the facts indicate that electrons of 
energy $\sim$120~TeV are distributed over 17~pc. 
If the acceleration site of the electrons is the pulsar, the
electrons should be transported over the distance within its
synchrotron lifetime. This condition requires that the
transportation velocity is at least $8.8\times10^{3}$~km~s$^{-1}$,
which may be explained by diffusion perpendicular to the magnetic field.}

\end{itemize}

\bigskip
We thank all Suzaku members.  HU is supported by 
Japan Society for the Promotion of Science (JSPS) 
Research Fellowship for Young Scientists. HM is supported by the MEXT,
Grant-in-Aid for Young Scientists~(B), 18740105, 2008, and
by The Sumitomo Foundation, Grant for Basic Science Research
Projects, 071251, 2007. This work was supported by the Grant-in-Aid 
for the Global COE Program "The Next Generation of Physics, Spun 
from Universality and Emergence" from the Ministry of Education, 
Culture, Sports, Science and Technology (MEXT) of Japan.

\end{document}